\documentclass[pre,twocolumn,aps,showpacs]{revtex4}
\usepackage{graphicx}
\usepackage{amsmath}
\usepackage{amsfonts}
\def\be{\begin{equation}}
\def\ba{\begin{eqnarray}}
\def\a{\alpha}
\def\b{\beta}

\def\d{\delta}
\def\D{\Delta}

\def\h{\eta}

\def\n{\nu}

\def\r{\rho}

\def\t{\tau}

\def\o{\omega}
\def\vo{\varpi}

\def\i{\int}

\def\tr{\text{tr}}
\def\Tr{\mbox{Tr}}

\def\ee#1{\label{#1}\end{equation}}
\def\ea#1{\label{#1}\end{eqnarray}}
\begin{document}
\title{Statistics of work performed on a forced quantum  
oscillator}
\author{Peter Talkner, P. Sekhar Burada, and Peter H\"anggi}
\affiliation{Institute of Physics, University of Augsburg,
D-86135 Augsburg, Germany}
\date{\today}
\begin{abstract}
Various aspects of the statistics of work performed by an external
classical force on a quantum
mechanical system are elucidated for a driven harmonic
oscillator. In this special case two parameters are introduced that 
are sufficient to completely characterize the force protocol. 
Explicit results for the
characteristic function of work and the respective probability
distribution are provided and discussed for three different types of
initial states of the oscillator: microcanonical, canonical and coherent
states.  Depending on the choice of the initial state the probability
distributions of the performed work may grossly differ. This result in
particular holds also true for
identical force protocols. General fluctuation and work theorems holding for
microcanonical and canonical initial states are confirmed.
\end{abstract}
\maketitle
\section{Introduction}
During the last decade various fluctuation and work theorems
\cite{fwt,ECM} have been formulated and discussed. They inter
alia characterize
the full nonlinear response of a system under the action of a time
dependent force \cite{J,J07}.
These theorems have been derived and experimentally
confirmed primarily for classical systems \cite{ex,BLR,BHSB}. 
Quantum mechanical generalizations
were proposed recently \cite{qm,M,RM,EM,TLH,TH,TMH,DL}. 

Conceptual problems though arise 
in the context of quantum mechanics 
if one tries to generalize those classical
relations that require for example the specification of a system's trajectory
extending over  
some interval of time, or the simultaneous measurement of
noncommuting observables.
For example, the measurement of work performed by an external force on
an otherwise isolated system may be accomplished in the framework of
classical physics in principle in two different ways. The first method
is based on two measurements of the energy, one at the beginning and
the second at the end of the considered process. This method becomes
unreliable in practice if the system is large and the work performed on the
system is negligibly small compared to the total energy of the system.
Such a situation typically arises if the system of interest, on which the
force exclusively acts, interacts with its environment. In order to
retain an isolated system, the large system made of the open system and
its environment must be considered. Again, the work performed on the
system results as the difference of the energies of the total system,
which both may be very large. 

For classical systems, this
unfortunate situation can be circumvented by a second method,  
by monitoring the state of
the relevant small system during the time when the force is acting. 
Having this information at hand one can  
determine the work by 
integrating the power supplied to the system at each instant of time. 
The respective power can be inferred from the registered state of
the system and the known force protocol. In a quantum
system a continuous measurement of even a single observable would strongly
influence and possibly manifestly distort the system's dynamics.
Apparently, only the first method of two energy measurements is
feasible, at least in principle, in the quantum context.  

An alternative method based on a continuous monitoring 
has recently
been suggested by Esposito and Mukamel \cite{EM} for open quantum
systems described
by Markovian quantum master equations. 
There the dynamics of the density matrix is 
mapped onto a classical rate process for which known fluctuation
theorems can be applied \cite{Seif}. This provides an interesting
formal approach but its physical meaning has remained unclear
\cite{EM}. 
Moreover,
this approach is restricted to open systems that only weakly
interact with their respective environments. 

In the present paper the distribution of work is discussed for the
exactly solvable system of a driven harmonic oscillator \cite{DL,H}. In
this case, 
the distribution of work is discrete. We provide formal expressions
for this distribution and its corresponding characteristic function
which are valid for all initial states of the system as well as for
all possible kinds of force protocols. In particular, we determine the
characteristic functions and distributions of the work for
microcanonical, canonical and coherent initial states which lead to
qualitatively different work distributions.

The paper is organized as follows. In Sect.~\ref{II} we review the general form
of the characteristic function of work performed on a system in terms
of a correlation function of the exponentiated Hamiltonians at the
initial and final time of the force protocol. We prove that this
particular expression indeed always represents a characteristic
function, i.e. the Fourier transform of a probability density.
Sect.~\ref{III} presents various fluctuation and work theorems for
canonical and microcanonical initial states. In Sect.~\ref{IV} general
expressions for the characteristic function and the corresponding
probability distribution of work are derived for a driven harmonic
oscillator. Moreover, the expressions for the first four cumulants are
derived. The dependence of the work distribution on the force protocol 
for microcanonical,
canonical and coherent initial states as well as its dependence on the
specific parameters of these initial states are investigated.


\section{Characteristic function of work} 
\label{II}
The response of a quantum system on a perturbation by a classical,
external force can be characterized by the change of energies
contained in the total system. The energy as an observable coincides
with its Hamiltonian $H(t)$  of
the total system. It includes
the external force and therefore depends on time. We will consider the
dynamics of the system only within a finite window of time $[t_0,t_f]$
during which the force is acting in a prescribed way, resulting in a
protocol of Hamiltonians which is denoted by $\{H(t)\}_{t_f,t_0}$.
Apart from the action of the external force the system is assumed to
be closed. Its dynamics is consequently governed
by a unitary time evolution $U_{t,t_0}$,
which is the solution of the Sch\"odinger equation
\be
\begin{split}
i \hbar \partial U_{t,t_0} / \partial t &= H(t) U_{t,t_0}, \\
U(t,t_0)&=1. 
\end{split}
\ee{SE}
As explained in the introduction, the work $w$ is measured as the
difference of the energies of the system at the final and initial times
$t_{f}$ and $t_{0}$. In a single measurement  the work is given by
the difference of two eigenvalues $e_{n}(t_{f})$ and $e_{m}(t_{0})$ of the
Hamiltonians $H(t)$ at the respective times $t_{f}$ and $t_{0}$,
i.e. by $w=e_{n}(t_{f})-e_{m}(t_{0})$. The inherent randomness of the
outcome of a quantum measurement in general leads to a  measured work
that is random.   
A complete description of the statistical properties of the work
performed on the system is provided by the characteristic function
$G_{t_{0},t_{f}}(u)$ 
which presents the Fourier transform of the probability density of the
work $p_{t_{f},t_{0}}(w)$, i.e.
\be
G_{t_{0},t_{f}}(u) = \int dw\: e^{iuw} p_{t_{f},t_{0}}(w).
\ee{Gp}
It can be expressed as a quantum correlation function of the
exponentiated Hamiltonian at the initial and the final time
\cite{TMH}, i.e.  
\be
\begin{split}
G_{t_{0},t_{f}}(u) &= \langle e^{iu H(t_{f})} e^{-iu H(t_{0})} \rangle
\\ 
& \equiv \Tr e^{i u H_{H}(t_{f})} e^{-iu H(t_{0})} \bar{\r}(t_{0}),
\end{split}
\ee{GC}
where 
\be
H_{H}(t_{f}) = U^{+}_{t_{f},t_{0}} H(t_{f}) U_{t_{f},t_{0}}
\ee{HH}
denotes the Hamiltonian in the Heisenberg picture. The density
matrix $\bar{\r}(t_{0})$ from the initial density matrix $\r(t_{0})$
as a result of the
measurement of the Hamiltonian $H(t_{0})$. It is given by
\be
\bar{\r}(t_{0}) = \sum_{n} P_{n}(t_{0}) \r(t_{0}) P_{n}(t_{0}),    
\ee{br}
where the operators $P_{n}(t_{0})$ denote the eigenprojection operators of the
Hamiltonian at time $t_{0}$, which present a partition of the unity
\be
\sum_{k} P_{n}(t_{0}) = 1.
\ee{P1}  
Before we apply the general expression (\ref{GC}) 
to a particular system and investigate its dependence on the
initial state $\r(t_{0})$, we discuss three general properties of the
correlation expression (\ref{GC}) which guarantee that the resulting
function $G_{t_{f},t_{0}}(u)$ indeed always presents a proper characteristic
function of a classical random variable $w$. This is the consequence
of the three following properties: \\ 
(i) $G_{t_f,t_{0}}(u)$ is a continuous function of $u$.\\
(ii)  $G_{t_f,t_0}(u)$ is a positive definite function of $u$, i.e. 
for all integer numbers $n$, 
all real sequences $u_1,u_2,\ldots u_n$, and all complex numbers
$z_{i}$, $i=1,2 \ldots n$ 
\be
\sum_{i,i'}^n  G_{t_f,t_0}(u_i-u_{i'})
z^*_{i} z_{i'} \geq 0 
\ee{pd}
holds. Here, the asterisk $z_{i}^*$ denotes the complex conjugate of
$z_{i}$.\\
(iii) $G_{t_f,t_0}(0) = 1$\\
According to a theorem by Bochner \cite{Boch} the properties (i-iii)
are necessary and sufficient conditions in order that the function
$G_{t_f,t_0}(u)$ 
is the Fourier transform of the probability measure of a random
variable. In short, the first condition insures that, strictly speaking,
the function $G_{t_{f},t_{0}}(u)$ is the Fourier transform of a
measure, the second condition assures that this measure is positive
and the third condition that it is normalized.
Hence the correlation expression eq. (\ref{GC}) always
defines a proper characteristic function. For a
proof of the properties (i-iii) we refer the reader to the appendix \ref{AA}.

\section{Canonical and microcanonical initial states}
\label{III}
In  experiments an external force is often applied on a system that
initially is found in a 
thermodynamic equilibrium state. Depending on whether the system was
in weak contact with a heat bath or was totally isolated from its
environment, the 
initial state of the system is described either by a canonical or a 
microcanonical density matrix. For both situations fluctuation and
work theorems are known. We will shortly review these relations.
\subsection{Work and fluctuation theorems for canonical initial
  states} 

If the initial density matrix is canonical, i.e. 
if 
\be
\r(t_0) = Z^{-1}(t_0) \exp \{ -\b H(t_0) \},
\ee{cs} 
where 
\be
Z(t_0) = \Tr \exp \{ - \b H(t_0) \} = e^{-\b F(t_{0})}
\ee{Z} 
denotes the partition function and $F(t_{0})$ the free energy, 
then $[H(t_0),\r(t_0)]=0$
and the first measurement of
the energy leaves the density matrix unchanged, such that
$\bar{\r}(t_0) = \r(t_0)$. With eq. (\ref{GC}) this leads to the
characteristic function of work for a canonical initial state which was
derived in Ref.~\cite{TLH}.
In this case, $G_{t_f,t_0}(u)$ can be
continued to an analytic function of $u$ for all $0\leq \Im u \leq
\b$ \cite{TH}. For the particular value $u=i\b$ the characteristic function
yields the mean value of the exponentiated work, $\langle \exp\{-\b w\}
\rangle$ and the 
correlation function expression (\ref{GC}) simplifies to the ratio of
the partition functions at the times $t_f$ and $t_0$, resulting in the
Jarzynski work theorem
\be
\langle e^{-\b w} \rangle    = Z(t_f)/Z(t_0) = \exp\left \{-\b
  (F(t_{f})-F(t_{0})) \right \},
\ee{J}
where $Z(t_f) = \tr \exp \{-\b H(t_f) \} = \exp \{- \b F(t_{f}) \}$.

Within the domain of analyticity $\mathcal{S} = \{u | 0 \leq \Im u
\leq \b\}$ the characteristic functions for the original and the time reversed
protocol are related to each other by the following formula, cf. \cite{TH} 
\be
G_{t_f,t_0}(u) = \frac{Z(t_f)}{Z(t_0)} G_{t_0,t_f}(-u+i\b),
\ee{GTC} 
where $G_{t_0,t_f}(u)$ refers to processes under the time reversed
protocol $\{H(t)\}_{t_o,t_f}$ starting from the canonical
state $Z^{-1}(t_f) \exp \{ -\b H(t_f)\}$. An inverse Fourier
transform leads to the Tasaki-Crooks fluctuation theorem, which
relates the probability
densities  of work $p_{t_f,t_0}(w)$ for a given protocol to
the density of
the work $p_{t_0,t_f}(w)$ for the time reversed protocol. This theorem
explicitly reads \cite{TH}
\be
\frac{p_{t_f,t_0}(w)}{p_{t_0,t_f}(-w)} = \frac{Z(t_f)}{Z(t_0)} e^{\b
  w} = e^{-\b(F(t_{f})-F(t_{0})-w)}.
\ee{TC}  
\subsection{Fluctuation theorems for microcanonical initial states}
A system in a microcanonical state is described by the density matrix
\be
\r(t_{0}) = \vo_{E}^{-1}(t_{0}) \d (H(t_{0})-E),
\ee{rmc}
where 
\be
\vo_{E}(t_{0}) = \Tr\: \d (H(t_{0})-E) = \exp\left \{S(E,t_{0})/k_{B}
\right \}
\ee{oE}
denotes the density of states as a function of the energy $E$ of the
system. The density of states can be expressed in terms of the
entropy of the system $S_{E}(t_{0})$ provided the spectrum of the
system Hamiltonian is sufficiently dense such that the density of
states becomes a smooth function on a coarsened energy scale. 
The microcanonical density
matrix commutes with the Hamiltonian $H(t_{0})$. Consequently,
$\bar{\r}(t_{0})$ and $\r(t_{0})$ coincide. 

The microcanonical quantum Crooks theorem
assumes the form \cite{TMH}
\be
\begin{split}
\frac{p_{t_{f},t_{0}}(E,w)}{p_{t_{0},t_{f}}(E+w,-w)} &=
\frac{\vo_{E+w}(t_{f})}{\vo_{E}(t_{0})}\\
& =
\exp \left \{\frac{S(E+w,t_{f})-S(E,t_{0})}{k_{B}} \right \}.
\end{split}
\ee{mcC}
Analogous to the canonical case it relates the probability density
$p_{t_{0},t_{0}}(E,w)$ of work $w$,
for a system starting in a microcanonical state with energy $E$, to the
respective quantity for the time reversed process starting
at energy $E+w$. This quantum theorem is formally identical to the
respective classical theorem \cite{CBK}.     

From the microcanonical Crooks theorem the probability density
relating to the time reversed process can be eliminated to yield the
so-called entropy-from-work theorem \cite{TMH}, reading:
\be
\vo_{E_{f}}(t_{f}) = \i dw\: \vo_{E_f-w}(t_{0})
p_{t_{f},t_{0}}(E_{f}-w,w) \;.
\ee{efw}  
This theorem allows one to determine the unknown density of states of a system
with Hamiltonian $H(t_{f})$ from the known density of states of a reference
system $H(t_{0})$ by means of the statistics of the work that is
performed on the system in a process that leads from the reference
system to the final system with unknown density of states. In the case
of systems with a sufficiently smooth density of states
the respective entropy can be determined. For further details
see in Ref.~\cite{TMH}.   
\section{Driven harmonic oscillator}
\label{IV}
To illustrate these concepts we consider 
an example which allows the analytical construction of the
probability of work. Specifically we consider a harmonic oscillator 
on which a time
dependent force acts during a finite interval of time.  
Its time evolution is governed 
by the Hamiltonian
\be
H(t) = \hbar \o a^+ a + f^*(t) a + f(t) a^+,
\ee{hoH}
where $\o$ denotes the angular frequency, and $a^+$ and $a$ creation and
annihilation operators, respectively, which obey the usual commutation
relation, i.e. $[a,a^+]=1$. The complex driving force $f(t)$ allows
for a coupling
to position and/or momentum of the oscillator.  
We assume that $f(t)$ vanishes for times $t\leq t_0=0$. 
It is our aim to study the influence of the initial state
$\r(t_{0})$ on the statistics of work performed on the oscillator. 
The measurement of
$H(t_{0}) = \hbar \o a^{+} a$
at time $t_{0}=0$ then yields the result $\hbar \o n$ with probability 
\be
p_{n} = \langle n|\r(t_{0})|n\rangle.
\ee{pn}
Accordingly, the oscillator is found in the state
\be
\bar{\r}(t_{0}) = \sum_{n} p_{n} |n\rangle \langle n|
\ee{bro}
immediately after this measurement. Putting this density matrix in the
general expression for the characteristic function, eq.~(\ref{GC}) one
obtains
\be
G_{t_f,t_{0}}(u) = \sum_n p_n e^{-iu\hbar \o
  n} \langle n |
e^{iu H_H(t_f)} |n \rangle\;.
\ee{G0} 
For the driven harmonic oscillator the diagonal matrix element of the
exponentiated Hamiltonian $H_{H}(t_{f})$ can be determined \cite{H}. For
details see the Appendix~\ref{B}. With the expression (\ref{nxy}) for
the matrix element $\langle n| \exp \left \{ iH_{H}(t_{f})\right \} |n
\rangle$ we
find
\begin{widetext}
\be
\begin{split}
G_{t_f,t_{0}}(u) =& \:  e^{iu |f(t_f)|^2/(\hbar \o)} 
\exp \left \{ \left ( e^{iu\hbar \o}-1
  \right )|z|^2 \right \}   \sum_{n=0}^\infty\sum_{k=0}^n p_n
\binom{n}{k} \frac{|z|^{2(n-k)}}{(n-k)!} e^{-iu\hbar \o (n-k)} \left
    (e^{iu\hbar \o}-1 \right )^{2(n-k)} \\
&=\:  e^{iu |f(t_f)|^2/(\hbar \o)} 
\exp \left \{ \left ( e^{iu\hbar \o}-1
  \right )|z|^2 \right \} \sum_{n=0}^{\infty}p_{n}L_{n}\!\left (4 |z|^{2}
\sin^{2}\frac{\hbar \o u}{2}\right )\;,
\end{split}
\ee{Gho} 
\end{widetext}
where $|f(t_f)|^2/(\hbar \o)$ denotes a uniform shift of the 
spectrum of the harmonic
oscillator due to the presence of the external force, cf.
eq. (\ref{L}), and 
\be
z = \frac{1}{\hbar \o} \i_0^{t_f} ds \dot{f}(s) \exp \{i \o s\} 
\ee{zf}
is a dimensionless functional of the driving force $f(t)$, cf. eq. (\ref{z}). 
This dimensionless quantity vanishes in particular for all-quasi
static forcings,
i.e. if the force changes only very slowly with 
$f(t)= g(t/t_{f})$ for $t_{f} \to \infty$,  
where $g(\t)$ is a continuously differentiable
function for $\t \in [0,1]$. We hence call $z(t)$ the {\it rapidity
parameter} of the force protocol.  
Finally, $L_{n}(x) =\sum_{k=0}^{n} \binom{n}{k} (-x)^{k}/k!$ denotes
the Laguerre polynomial of order $n$ \cite{RG}.

Introducing the cumulant generating function $K(u)= \ln G(u)$ one
obtains 
the cumulants of work $k_{n}$  
as the $n$th derivatives of $K(u)$  with respect to $u$ taken at $u=0$
\cite{vK}, i.e. $k_{n} = (-i)^{n} d^{n }K(0)/d u^{n}$. The 
first four cumulants become:
\begin{align}
k_{1} &= \langle w \rangle \nonumber\\ &= \frac{|f(t_f)|^2}{\hbar \o} +\hbar \o
|z|^2, \label{aw}\\
k_{2} &=\langle w^{2} \rangle -\langle w \rangle^{2}\nonumber\\
 &= 2 (\hbar \omega)^2 |z|^2
\left ( \langle a^+ a \rangle_0 +\frac{1}{2} \right ),\label{ww}\\
k_{3} &=\langle w^{3} \rangle- 3 \langle w^{2} \rangle \langle w
\rangle +2 \langle w
\rangle^{3}\nonumber\\ 
&= \left (\hbar \o \right )^{3} |z|^{2},\\
k_{4}&=  \langle w^{4} \rangle -4 \langle w^{3} \rangle\langle w
\rangle -3 \langle w^{2} \rangle^{2} \nonumber \\
& \quad +12 \langle w^{2} \rangle \langle
w \rangle^{2} - 6 \langle w \rangle^{4}\nonumber\\ 
&= \left (\hbar \o \right )^{4}
|z|^{2} \left \{ 1+4\langle a^{+}a \rangle_{0} + 6 \left [\langle
  a^{+}a(a^{+}a-1)\rangle_{0} \right . \right .\nonumber \\
&\left . \left .\quad - 2\langle a^{+}a \rangle_{0} \right ]|z|^{2}\right \}.   
\label{cu}
\end{align}
The odd cumulants of the work are independent of the
initial preparation. The even cumulants depend on the
factorial moments of the number operator $a^{+}a$ with respect to
the initial state $\bar{\r}(t_{0})$ such as $\langle
a^{+}a \rangle_{0} = \sum_{n} n p_{n}$ and $\langle
a^{+}a(a^{+}a -1) \rangle_{0} = \sum_{n} n(n-1) p_{n}$, where $p_{n}$
is defined in eq.~(\ref{pn}).  Moreover, all
cumulants apart from the first one 
vanish for forcings with $z=0$. This holds true in particular for all quasi-static
force characteristics. The underlying work probability density then
shrinks to a delta function at $w=|f(t_{f})|^{2}/(\hbar \o)$. 

In general, the work probability density follows from the characteristic
function by means of an inverse Fourier transformation. Rather than
the characteristic function itself we first consider the function
$\mathcal{G}(u)\equiv \exp \left \{ -iu|f(t_{f})|^{2}/(\hbar \o)\right\}
G_{t_{f},t_{0}}(u)$. Upon expanding $\exp\left \{|z|^{2} \exp\left
  \{iu\hbar \o\right \} \right \}$
into a series of powers of $|z|^{2}$ we obtain for $\mathcal{G}(u)$ a
Laurent series in the variable $\exp\left\{iu\hbar \o\right\}$. The
inverse Fourier transformation is given by a series of delta
functions $\d(w-\hbar \o r)$, with $r \in \mathbb{Z}$, with
weights
\be
\begin{split}
q_r =& e^{-|z|^2 } \sum_{m,n=0}^\infty \sum_{k=0}^n
\sum_{l=0}^{2k}(-1)^{2k-l} p_n  \\
&\quad\times \frac{|z|^{2(k+m)}}{m!\: k!}
\binom{n}{k} \binom{2k}{l} \d_{l+m,k+r}\\
=& e^{-|z|^2}\sum_{n=0}^\infty \sum_{k=0}^n \sum_{l=0}^{\text{min}
  (k+r,2k)}
(-1)^{2k-l}  p_n  \\
&\quad \times \frac{|z|^{2(2k+r-l)}}{(k+r-l)!\:k!} \binom{n}{k}
\binom{2k}{l}\;. 
\end{split}
\ee{q}    
The factor  $\exp \left \{ -iu|f(t_{f})|^{2}/(\hbar \o)\right\}$, by
which $\mathcal{G}(u)$ has to be multiplied to yield
$G_{t_{f},t_{0}}(u)$, gives
rise to a constant shift such that the probability density of work
performed on a harmonic oscillator assumes the result 
\be
p_{t_f,0}(w) = \sum_r q_{r} \:\d\!\left (w-(\hbar \o r
+\frac{|f(t_{f})|^{2}}{\hbar \o})\right ) \;.
\ee{pt0}
In the next Section we will investigate the influence of the initial
state on the statistics of the work.
\subsection{Distributions of work for different initial states}
As particular examples of initial states we will discuss
microcanonical, canonical and coherent states.\\
\subsubsection{Microcanonical initial state}
\begin{figure}
\includegraphics[width=8cm]{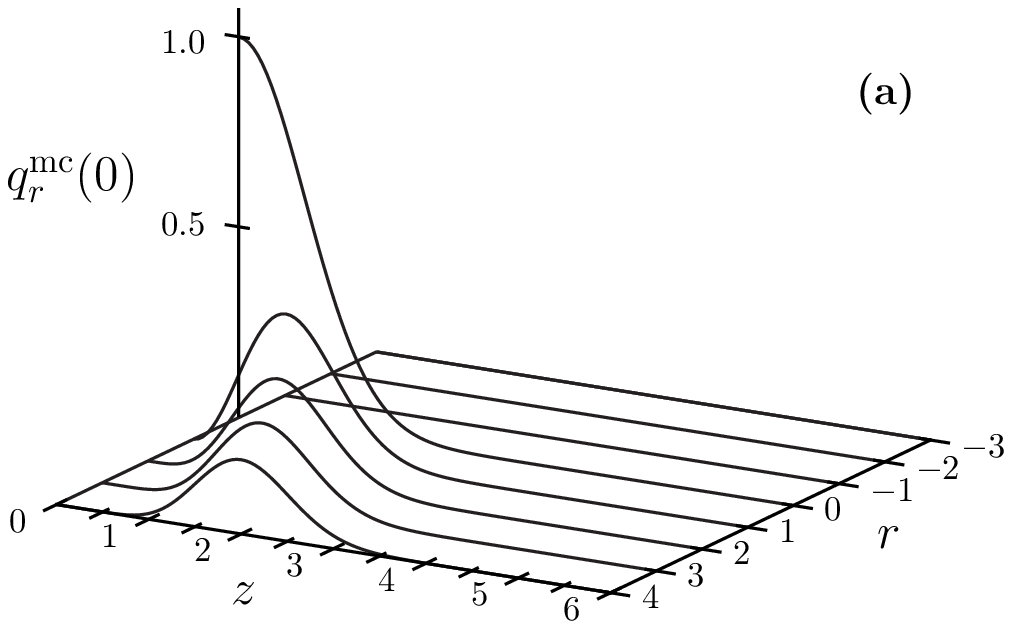}\hfill 
\includegraphics[width=8cm]{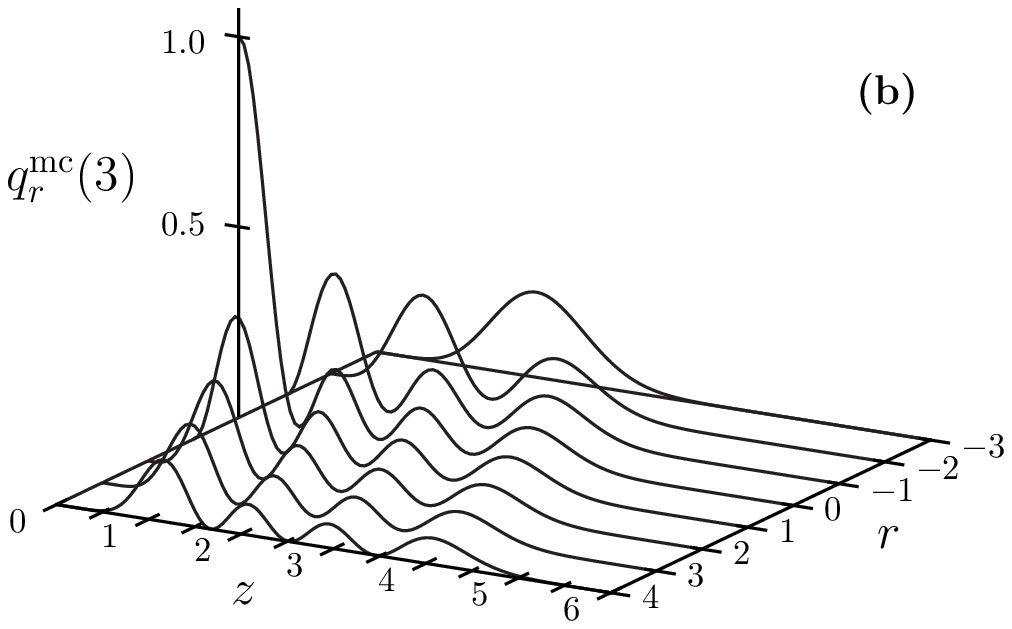} 
\caption{The probabilities $q^{\text{mc}}_{r}(n_{0})$ for two
microcanonical initial states with $n_{0}=0$ (a) and $n_{0}=3$ (b) are 
depicted for $r=-3
\ldots 4$, as functions of the rapidity parameter $z$ in
eq.~(\ref{z}).  In both cases the
distribution collapses at $r=0$ for adiabatic forcing corresponding to
$|z|=0$ and broadens with increasing $|z|$. Obviously, when starting in
the ground state the oscillator cannot deliver work whence the
probability for negative $r$ strictly vanishes. ``Stimulated
emission'' becomes possible from an excited state at finite driving
rapidity $z$ leading to nonzero probabilities
$q^{\text{mc}}_{r}(n_{0})$ at negative values of $r$ in panel (b).}      
\label{fz}
\end{figure}

\begin{figure}
\includegraphics[width=8cm]{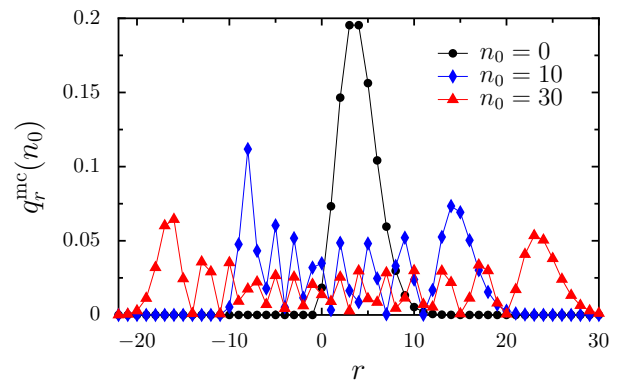}
\caption{(Color online) The probabilities $q^{\text{mc}}_{r}(n_{0})$ for a
microcanonical initial state with $n_{0}=0$ (circles) and $n_{0}=10$
(diamonds) and $n_{0}=30$ (crosses) are
compared for a fixed rapidity parameter $|z|=2$ and $r=-22 \ldots
30$. The lines serve as a guide for the eye.}
\label{fn}
\end{figure}

For a microcanonical initial state with energy $\hbar \o n_{0}$ the
density matrix becomes
\be
\r(t_{0}) = \bar{\r}(t_{0}) = |n_{0}\rangle \langle n_{0} |.
\ee{mco}
The characteristic function then reads
\be
\begin{split}
G^{\text{mc}}_{t_{f},t_{0}}(n_{0},u) &= e^{iu|f(t_{f})|^{2}/(\hbar
  \o)} \exp\left 
  \{ \left( e^{iu\hbar \o} -1\right)|z|^{2}\right \} \\
&\quad \times L_{n_{0}}\!\left ( 4|z|^{2} \sin^{2} \frac{\hbar \o u}{2}\right )
\end{split}
\ee{mccf0}   
and, accordingly, the probability $q^{\text{mc}}_{r}(n_{0})$ to find a change of
energy by $w=\hbar \o r +|f(t_{f})|^{2}/(\hbar \o)$ emerges as 
\be
\begin{split}
q^{\text{mc}}_{r}(n_{0})& = e^{-|z|^{2}} \sum_{k=0}^{n_{0}} \sum_{l=0}^{\min
  (k+r,2k)}\frac{(-1)^{2k-l}}{(k+r-l)!\: k!}\\
&\quad \times \binom{n}{k} \binom{2k}{l}
|z|^{2(2k+r-l)}.
\end{split}
\ee{mcq}
As expected from the behavior of the moments, all probabilities
$q^{\text{mc}}_{r}(n_{0})$ with $r\neq 0$ vanish for quasi-static forcing, i.e. if
$z \to 0$. The dependence of $q^{\text{mc}}_{r}(n_{0})$ for $n_{0}=0$ and $3$ 
as well as for the eight lowest values of
$r$ on the parameter $z$ is displayed in Fig.~\ref{fz}. With
increasing values of the rapidity parameter $z$ the distribution is
becoming broader.
For the fixed value of $|z|=2$ the distribution
$q^{\text{mc}}_{r}(n_{0})$ is compared for the three initial states  with
$n_{0}=0$ , $10$ and $30$ in
Fig.~\ref{fn}. 
With increasing value of $n_{0}$ the distributions
become broader. They develop a slightly asymmetric shape with 
higher peaks at negative values of $r$ compared to those at positive
$r$ values. Between these dominant peaks the probability still
displays pronounced variations.\\
For a harmonic oscillator, the microcanonical Crooks theorem reduces
to the relation $q^{\text{mc}}_{r}(n) = q^{\text{mc}}_{-r}(n+r)$. One can show that
this symmetry is fulfilled by the probabilities $q^{\text{mc}}_{r}(n)$
given by eq.~(\ref{mcq}).
\subsubsection{Canonical initial state}
For a canonical density matrix 
\be
\r(t_{0}) = (1-e^{-\b \hbar \o }) e^{-\b \hbar \o a^{+} a}
\ee{cdm}
 the initial states are
distributed according to $p_{n} = e^{-\b
  \hbar \o n}/(1- e^{-\b \hbar \o})$. This allows one to perform the
sum over $n$ in the expression for the characteristic function
(\ref{Gho}) in
terms of the generating function of the Laguerre polynomials, cf. \cite{RG}
yielding the expression
\begin{widetext}
\be
G^{\text{c}}_{t_{f},t_{0}}(\b,u) = \exp \left \{ \frac{iu
    |f(t_{f})|^{2}}{\hbar \o} +\left(e^{iu\hbar
      \o}-1 \right )|z|^{2} - 4|z|^{2} \frac{\sin^{2}(\hbar \o u/2)}{e^{\b \hbar
      \o}-1}   \right \}. 
\ee{Gca}
\end{widetext}
Putting $u=i\b$ one finds that the two terms in the
exponent which are proportional to $|z|^{2}$ cancel each other, such
that one obtains 
\be
\langle e^{\b w} \rangle = G^{\text{c}}_{t_{f},t_{0}}(\b,i\b) =
\exp\left \{-\b
  |f(t_{f}|^{2}/(\hbar \o)\right \}. 
\ee{Jho}
 The free energy difference of two oscillators with Hamiltonians
 $H(t_{0}) =\hbar \o a^{+}a$ and  $H(t_{f}) =\hbar \o a^{+}a
 +f^{*}(t_{f}) a +f(t) a^{+})$ each one staying in a canonical state at the
  temperature $\b$ is given by $\D F = F(t_{f}) - F(t_{0}) =
 |f(t_{f})|^{2}/(\hbar \o)$ in accordance with Jarzynski's work theorem.

The probability $q^{\text{c}}_{r}(\tilde{\b})$ to find the work $w= \hbar \o r
+|f(t_{f})|^{2}/(\hbar \o)$ if the system starts in a canonical state becomes
\be
\begin{split}
q^{\text{c}}_{r}(\tilde{\b})& = e^{-|z|^{2}} \left (1 - e^{-\tilde{\b}
  } \right )\sum_{n=0}^{\infty} \sum_{k=0}^{n} \sum_{l=0}^{\min
  (k+r,2k)} 
  (-1)^{l} e^{-\tilde{\b}  n} \\
&\quad \times \frac{|z|^{2(2k+r-l)}}{(k+r-l)!
    k!} \binom{n}{k}
\binom{2k}{l}
\end{split}
\ee{qc}
where $\tilde{\b} = \b \hbar \o$ denotes the inverse dimensionless
temperature of the initial state.
The expression for $q^{c}_{r}(\tilde{\b})$ can be further simplified to read
\be
q^{\text{c}}_{r}(\tilde{\b})= e^{-|z|^{2} \coth
  (\tilde{\b}/2)} e^{\tilde{\b}r/2} I_{r}\left (\frac{|z|^{2}}{\sinh
    \tilde{\b}/2} \right )
\ee{qc2}
where $I_{\n}(x)$ denotes the modified Bessel function of first kind of
order $\n$ \cite{RG}. For details of the derivation see Appendix~\ref{C}.

\begin{figure}
\includegraphics[width=8cm]{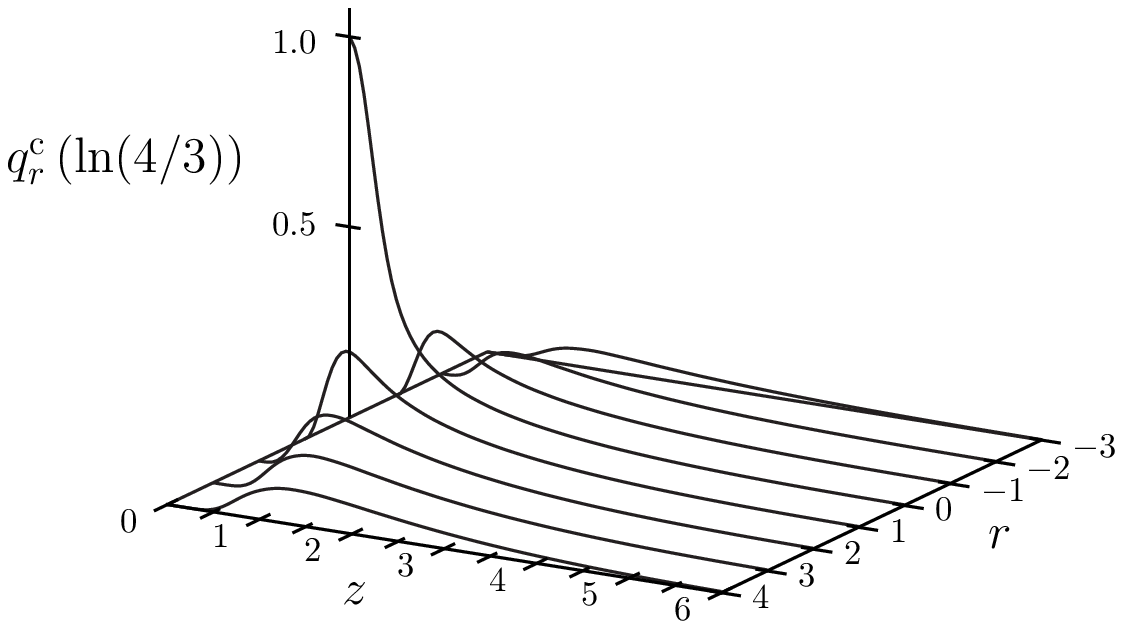}
\caption{The probabilities $q^{\text{c}}_{r}(\tilde{\b}=\ln(4/3))$ for a
canonical initial state are displayed for $r=-3
\ldots 4$ as a function of the parameter $z$. For the sake of
comparability the dimensionless
inverse temperature is chosen such that the average energy in the
initial state coincides with the energy $3 \hbar \omega$ 
of the microcanonical state in Fig.~\ref{fz} (b).   }
\label{fcz}
\end{figure}

\begin{figure}
\includegraphics[width=8cm]{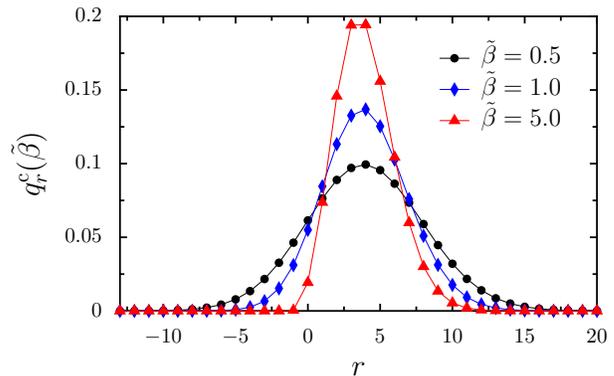}
\caption{(Color online) The probabilities $q^{\text{c}}_{r}(\tilde{\b})$ for a
canonical initial state are displayed as functions of  $r$ 
for  $|z|=2$ and different values of the dimensionless inverse
temperature $\tilde{\b} = 0.5$ (boxes), $1$ (circles) and $5$
(crosses). The lines serve as a guide for the eye.}  
\label{fcb}
\end{figure}

Note that the following detailed balance like symmetry relation exists,
\be
q^{\text{c}}_{-r}(\tilde{\b}) = e^{-\tilde{\b}r} q^{\text{c}}_{r}(\tilde{\b}),
\ee{qrmr}
relating the occurence of 
positive and negative work.   
In Fig.~\ref{fcz} the $z$ dependence of $q^{c}_{r}(\tilde{\b})$ for
$\tilde{\b}=\ln(4/3)$ is compared for a few small values of $r$. One finds
that due to the average over the canonical initial distribution 
the multipeaked structure of the microcanonical distribution as a
function of the rapidity parameter $|z|$ disappears and only
a single peak remains for each value of $r$.
The temperature dependence of the work distribution is illustrated
in Fig.~\ref{fcb}.
Finally, we verify the validity of the Tasaki-Crooks theorem (\ref{TC})
for a driven oscillator. For this purpose we
consider the probability density $p_{t_{0},t_{f}}(-w)$ for the time
reversed protocol. Since the absolute values of the rapidity parameters
coincide for the original and the time reversed protocol the
probability density of work for the reversed protocols becomes
\be
p_{t_{0},t_{f}}(-w) = \sum_{r=-\infty}^{\infty} q_{r} \d\left (-w-(\hbar \o r
-\frac{|f(t_{f})|^{2}}{\hbar \o}) \right)\; ,
\ee{p0f}
where we took into account the overall shift of the spectrum by the
reversed protocol as
$-|f(t_{f})|/(\hbar \o)$.
Multiplying both sides of eq.~(\ref{p0f}) with $\exp \left \{ -\b( \D
  F -w) \right \}=\exp \left \{ -\b( |f(t_{f})|^{2}/(\hbar \o) -w)
\right \} $  one obtains 
\be
\begin{split}
e^{  -\b( \D
  F -w) } p_{t_{0},t_{f}}(-w) &= \sum_{r} e^{ -\b
  \left (
  |f(t_{f})|^{2}/(\hbar \o)  -w \right )}
q^{\text{c}}_{r}(\tilde{\b}) \\
&\quad \times \d\left (-w-(\hbar \o r 
-\frac{|f(t_{f})|^{2}}{\hbar \o}) \right)\\
&=\sum_{r} e^{\tilde{\b}r} q^{c}_{r}(\b)\\
&\quad \times \d \left (w +(\hbar
  \o r -\frac{|f(t_{f})|^{2}}{\hbar \o}) \right )\\
&= p_{t_{f},t_{0}}(w),
\end{split}
\ee{TCho}
in accordance with the Tasaki-Crooks theorem (\ref{TC}).

\subsubsection{Coherent initial state}
An oscillator prepared in a coherent state $|\a \rangle$ is described
by the density matrix
\be
\r(t_{0}) = |\alpha \rangle \langle \alpha |
\ee{rcs}
where 
\be
|\a \rangle = e^{\a a^{+} + \a a} |0 \rangle
\ee{acs}
and $|0 \rangle $ is the normalized ground state of the oscillator
satisfying $a |0 \rangle =0$.
Note that the coherent state density matrix does {\it not} commute with the
Hamiltonian $H(t_{0})$. The measurement of $H(t_{0})$ modifies the
coherent state (\ref{rcs})  by
projecting it onto the eigenstates $|n \rangle = (a^{+})^{n}
/\sqrt{n!} |0 \rangle$ of this Hamiltonian leading to 
\be
\bar{\r}(t_{0}) = e^{-|\a|^{2}} \sum_{n} \frac{|\a|^{2n}}{n!} |n
\rangle \langle n |
\ee{brcs}
This implies a Poissonian distribution of the respective  
energy eigenvalues $\hbar \o n$ 
\be
p^{\text{cs}}_{n} = \frac{|\a|^{2n}}{n!} e^{-|\a|^{2}}, 
\ee{pcs}
which yields for the characteristic function of work (\ref{Gho}) a
closed expression of the form
\be
\begin{split}
G^{\text{cs}}_{t_{f},t_{0}}(\a,u) &= \exp\left
  \{\frac{iu|f(t_{f})|^{2}}{\hbar \o} +|z|^{2} \left(e^{i\hbar \o u}-1
  \right ) \right \}\\ 
& \quad \times J_{0}\left (4\left |\a z \sin 
  \frac{\hbar \o u}{2} \right | \right ) 
\end{split}
\ee{Gcs}  
where $J_{0}(x)$ is the Bessel function of order zero, cf. Ref.~\cite{RG}.
For the probability $q^{\text{cs}}_{r}(\a)$ of work one obtains with
eq.~(\ref{q})
\begin{widetext}
\be
q^{\text{cs}}_{r}(\a) = \;e^{-|z|^{2}} \sum_{m=0}^{\infty} 
  \frac{|z|^{2m} \:|\a z|^{2|m-r|}}{m! \left( |m-r|!\right)^{2}}
  {}_{1}F_{2}
  \left ( |m-r|+\frac{1}{2};|m-r|+1,2|m-r|+1;-4|\a z|^{2}\right )
\ee{qcs}
\end{widetext}
where $_{1}F_{2}(a;b,x;x)$ denotes a generalized hypergeometric
function \cite{RG}. For details of the derivation see the Appendix~\ref{D}. 
 
\begin{figure}
\includegraphics[width=8cm]{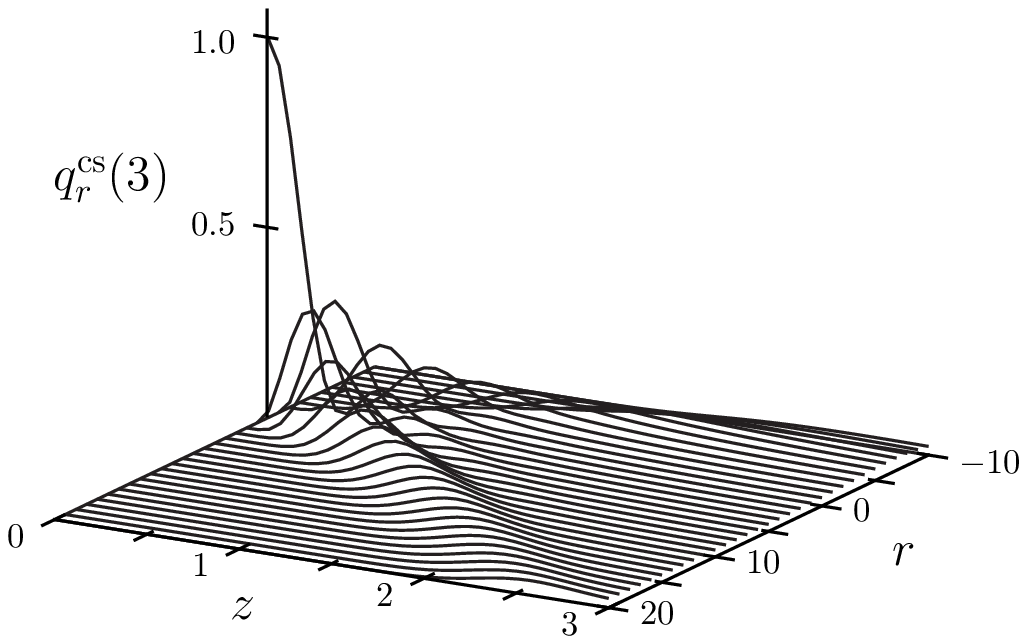}
\caption{The probabilities $q^{\text{cs}}_{r}(\a)$ for a
coherent state with parameter $|\a|=3$ are displayed 
for $r = -10 \ldots 20$ as functions of  $z$.}  
\label{fcsz}
\end{figure}

The dependence of  the probabilities $q^{\text{cs}}_{r}(\a)$ on the
rapidity parameter $|z|$ is illustrated in Fig.~\ref{fcsz} for $r$
values ranging from $-10$ to $20$. 
\begin{figure}
\includegraphics[width=8cm]{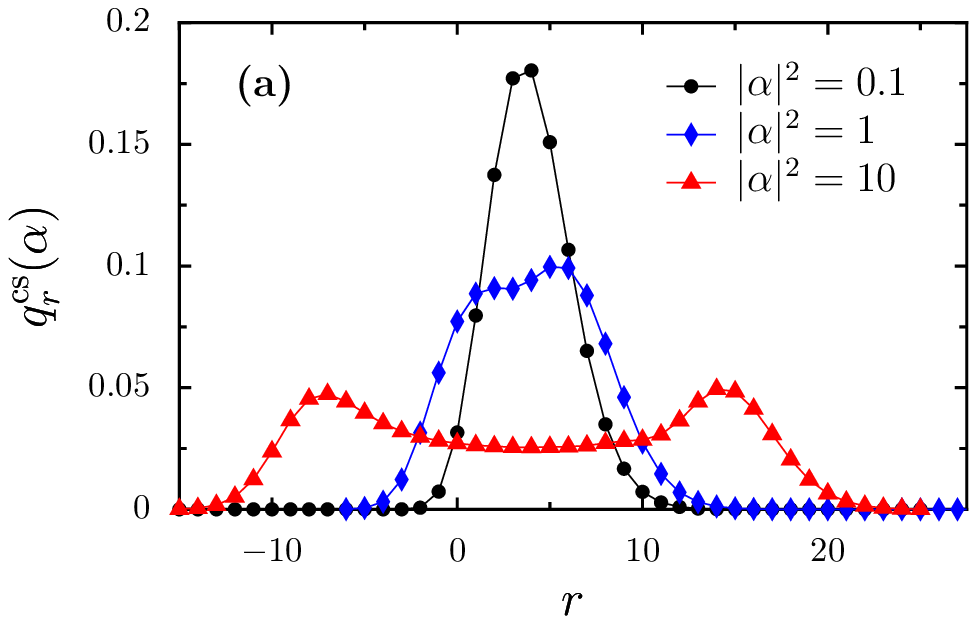}
\hfill
\includegraphics[width=8cm]{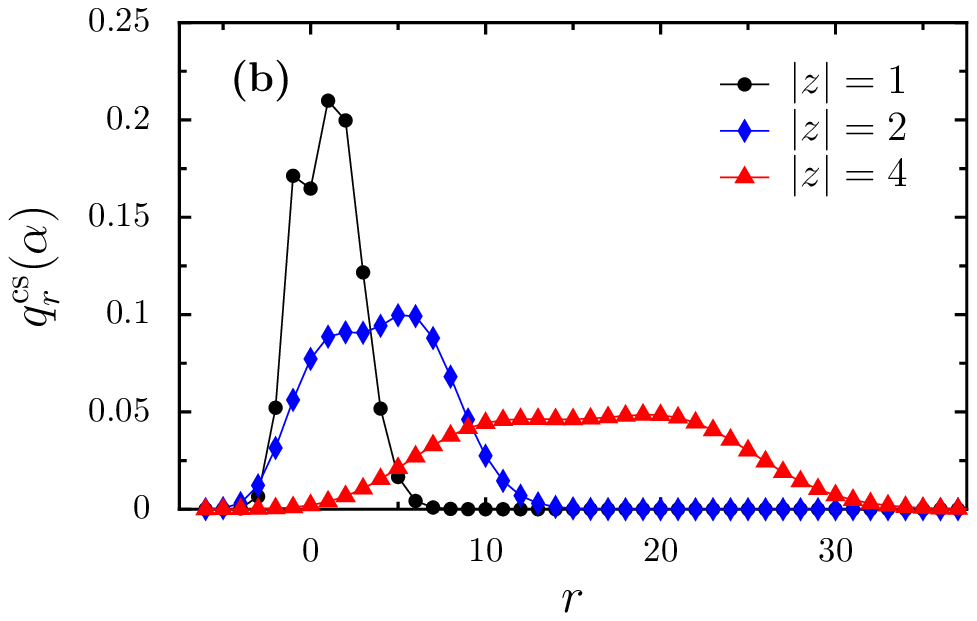}
\caption{(Color online) The distribution of work performed on an
  oscillator which 
  initially is prepared in a coherent state $|\a \rangle$ for
  different values of $\a$ in panel (a) and of the rapidity parameter
  $z$ in panel (b).
In panel (a) the rapidity parameter has the value $|z|=2$. In panel
(b) the coherent state parameter has the value  
 $|\a|^{2} =1$. }
\label{f2}
\end{figure}
Increasing values of $z$ lead to a broadening of the distribution
and also to a shift towards larger values of $r$, see also panel (b)
of Fig.~\ref{f2}.
This is in accordance with eq. (\ref{aw}) and (\ref{ww})
for the first two cumulants of the work, which both increase with
$|z|^{2}$.
Panel (a) of Fig.~\ref{f2} shows the dependence of the probabilities
$q^{\text{cs}}_{r}(\a)$ on the parameters $\a$. Increasing
$\a$ also leads to a broadening of the work
distribution without influencing its mean value, cf. also eq. (\ref{aw}). 


\begin{figure}
\includegraphics[width=8cm]{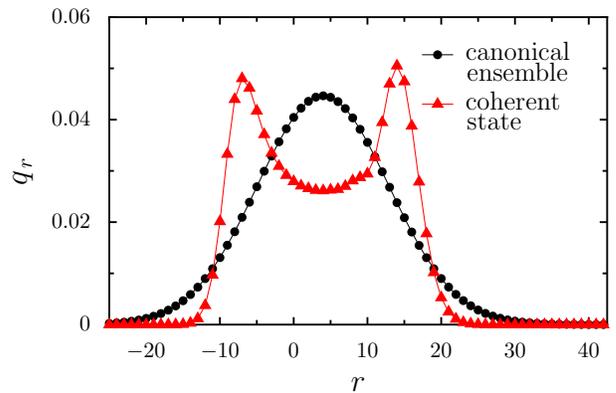}
\caption{(color online) The distribution of work is compared for a
  canonical and a  
  coherent initial state subject to the same force protocol with
  rapidity parameter $z=2$. With $\b \hbar \o  =0.1$ and  $|\a|^2\approx
  9.51$ the expectation values of the energies agree in the two initial
  states such that according to eqs.~(\ref{aw}) and (\ref{ww}) the
  first and second moments of the work also coincide. Still the
  distributions of work grossly differ from each other. }
\label{f3}
\end{figure}  
In Fig.~\ref{f3} the probabilities $q_r$ are depicted for different
initial states. In one case the oscillator is initially prepared in a
canonical state at inverse dimensionless temperature $\tilde{\b}=\b
\hbar \o =0.1$. 
In the other case, the oscillator stays in a coherent state
$|\a\rangle$, where the absolute value of $|\a|$ is chosen such that
the mean excitation number is the same for both states, i.e. $|\a|^2 =
\exp\{ -\b \hbar \o\} / (1- \exp\{ -\b \hbar \o\} )$. For $\b \hbar \o
=0.1$ one finds $|\a|^2 \approx 9.51$. The two
oscillators then are subjected to protocols with the same
rapidity parameter 
$|z|=2$. According to eqs. (\ref{aw}) and (\ref{ww}) the first two
moments of the work performed on the oscillator coincide. Yet the
distribution of weight factors $q^{\text{c}}_r(\tilde{\b})$ and
$q^{\text{cs}}_r(\a)$ distinctly differ. Whereas the
distribution is pronouncedly bimodal in case of the coherent state, it
is unimodal for the canonical state. The weight factors
$q^{\text{c}}_r(\tilde{\b})$ almost 
perfectly fall onto the Gaussian probability density which has the
same first two
moments as the discrete distribution given by $q_r$.

\section{Conclusions}
In this work we studied the statistics of work performed on an
externally driven quantum mechanical oscillator by means of a
correlation function expression of the work. We demonstrated that this
particular expression indeed always represents a proper characteristic
function of a random variable, which is the performed work in the present
context. The proof given here is based on Bochner's theorem. It holds
for general 
quantum mechanical systems, not only for harmonic oscillators.

The considered force linearly couples to the position and momentum of
the oscillator. It may describe the influence of an electric
field on charged particles in a parabolic trap or the external forcing
of a single electromagnetic cavity mode. For this
type of additive forcing, the frequency of the oscillator remains unchanged and
therefore the level spacing of the eigenvalues of the Hamiltonian is
not influenced by the force. The spectrum is only shifted as a whole.
As a consequence the work performed on the oscillator is, as a positive
or negative integer of the level spacing, 
a discrete random variable. We determined the first few cumulents of
the work for arbitrary force protocols and initial states. A
complementary study for a parametrically forced oscillator was
recently 
performed by Deffner and Lutz \cite{DL}.     

It turns out that for the harmonic oscillator 
the statistics of work depends on the force
protocol $\{f(t)\}_{t_{f},t_{0}}$ only through two real parameters, 
which are (i) the shift of the spectrum, given by $L(t_{f})=
|f(t_{f})|^{2}/(\hbar \o)$, 
and (ii) the absolute value of the dimensionless quantity $z =
\i_{t_{0}}^{t_{f}}\dot{f}(s) \exp \{i \o s \}$. 
This parameter vanishes for all quasi-static processes
and therefore presents a measure of the rapidity of the force
protocol.
While the presence of $L(t_{f})$ only causes an overall shift of the
possible values of the work, the rapidity parameter $|z|$ also
influences its distribution. Typically, the distributions move towards 
larger values of work $w$ and become broader with increasing rapidity $|z|$.

We also demonstrated that different initial states of the system such
as microcanonical, canonical or coherent states, have a large influence on
the work statistics. We further note that two different initial
density matrices with the 
same diagonal elements with respect to the energy eigenbasis of the
Hamiltonian $H(t_{0})$ lead to identical work distributions
even though the two density matrices may be very different in other
respects. For example, the coherent pure state $|\a\rangle \langle \a|$
and the mixed state $\exp\{-|\a|^{2 n}\}\sum_{n} |\a|^{2}/n! |n\rangle
\langle n| $ cannot be distinguished by means of their respective work
statistics. This statistics is also insensitive to the phase of a
coherent state.  

\acknowledgments
This work has been supported by the Deutsche Forschungsgemeinschaft
via the Collaborative Research Centre SFB-486, project A10. Financial
support of the German Excellence Initiative via the {\it Nanosystems
  Initiative Munich} (NIM) is gratefulle acknowledged as well.

\appendix
\section{Proof of the properties of $G_{t_f,t_0}(u,v)$}
\label{AA}
We prove that the conditions of Bochner's theorem are fulfilled, and
consequently $G_{t_{f},t_{0}}(u)$ is a proper characteristic function.
  
Proof of property (i): $G_{t_{f},t_{0}}(u)$ is a continuous function
of $u$. The Hamiltonian operators at the two times of measurement
$t_{0}$ and $t_{f}$ are selfadjoint operators. According to the
theorem of Stone \cite{Yosida}, each of the exponential operators $\exp \left
  \{ -iu H(t_{0})\right \}$ and  $\exp \left
  \{ iu H_{H}(t_{f})\right \}$ forms a strongly continuous one-parameter 
group of unitary operators with parameter $u$. As the trace of a
product of two strongly continuous operator valued functions of $u$
with the density operator $\bar{\r}(t_{0})$, which is a trace class
operator and independent of $u$, the
characteristic function (\ref{GC}) is a continuous function of $u$.

Proof of property (ii): $G_{t_{f},t_{0}}(u)$ is a positive definite
function of $u$.
Using the cyclic invariance of the trace and the  fact that $H(t_{0})$
and $\bar{\r}(t_{0})$ commute with each other, we can rewrite the left
hand side of the inequality (\ref{pd}) as
\begin{widetext}
\be
\sum_{i,j}^{n} G_{t_{f},t_{0}}(u_{i}-u_{j}) z^{*}_{i} z_{j} =
\sum_{i,j}^{n} \Tr \:e^{i(u_{i}-u_{j})H_{H}(t_{f})} \:
e^{-i(u_{i}-u_{j})H(t_{0})} \bar{\r}(t_{0}) z_{i}^{*} z_{j}
= \Tr A^{+} A \:\bar{\r}(t_{0}) \geq 0\;,  
\ee{pos}   
\end{widetext}
where 
\be
A= \sum_{i}z_{i}^{n} e^{-iu_{i}H_{H}(t_{f})} e^{iu_{i} H(t_{0})}
\ee{A}
is a bounded operator and $A^{+}$ its adjoint. The last inequality in 
(\ref{pos}) immediately 
follows with the positivity of $A^{+}A$ and of the density matrix
$\bar{\r}(t_{0})$.   

Proof of property (iii): $G_{t_{f},t_{0}}(0) =1$. 
For $u=0$ the exponential operators $\exp \left
  \{ -iu H(t_{0})\right \}$ and  $\exp \left
  \{ iu H_{H}(t_{f})\right \}$ become unity. The trace over the
density matrix $\bar{\r}(t_{0})$ reduces by means of
eqs. (\ref{br}), (\ref{P1}) to the trace of the initial
density matrix $\r(t_{0})$, which is one.
\section{The matrix element $\langle n| \exp \{iu H_H(t_f)\} |n\rangle
  $}
\label{B} 
The total time rate of change of the Hamiltonian $H_H(t)$ coincides
with its partial derivative with respect to the time which for the
driven oscillator becomes, cf. eq. (\ref{hoH}),
\be
\frac{d H_H(t)}{dt} = \dot{f}^* (t) a_H(t) - \dot{f}(t) a^+_H(t),
\ee{dH}
where $a_H(t)$ and $a^+_H(t)$ denote annihilation and creation
operators, respectively, in the Heisenberg picture, which are given by
\be
\begin{split}
a_H(t) &=   e^{-i \o t} a -\frac{i}{\hbar} \i_0^t ds e^{-i\o(t-s)} f(s)\\
a^+_H(t) &=  e^{i \o t} a^+ +\frac{i}{\hbar} \i_0^t ds e^{i\o(t-s)} f^*(s)  
\end{split}
\ee{aH}
This yields for $H_H(t_f)$
\be
H_H(t_f) = \hbar \o a^+ a + B^*(t_{f})a + B(t_{f}) a^+ + C(t),
\ee{HtH}
where
\be
\begin{split}
B(t_{f}) &= \i_0^{t_{f}} ds \dot{f}(s) e^{i \o s}\\
C(t_{f})& = \frac{i}{\hbar} \i_0^{t_{f}} ds \i_0^s ds' \left [ \dot{f}(s)
  f^*(s') e^{i\o(s-s')} \right .\\
&\quad \left . - \dot{f}^*(s) f(s') e^{-i\o (s-s')} \right ].
\end{split}
\ee{FG}
The unitary operator
\be
V= e^{z a^+ -z^* a}
\ee{V}
with 
\be
z = \frac{B(t_{f})}{\hbar \o}
\ee{z}
transforms $H_H(t)$ into
\be
V H_H(t_{f}) V^+ = \hbar \o a^+ a + L(t_{f}),
\ee{VHV}
where 
\be
L(t_{f}) = C(t_{f})-\frac{|B(t_{f})|^2}{\hbar \o}
=\frac{|f(t_{f})|^2}{\hbar \o} .
\ee{L}
Note that $V$ induces a shift of the creation and annihilation
operators
\be
\begin{split}
V a V^+ = a - z, \quad
V a^+ V^+ = a^+ - z^*
\end{split}
\ee{Va}
and further note that, when acting on the groundstate $|0\rangle$ with
$a|0\rangle =0$, the operator $V$ yields the coherent state $|z\rangle$, i.e.
\be
V|0\rangle = | z \rangle.
\ee{V0}
One finds with these properties
\begin{align}{\allowdisplaybreaks}
\langle n | e^{iu H_H(t_{f})}|n\rangle &= \frac{1}{n!} \langle
z|(a-z)^n \nonumber \allowdisplaybreaks\\
&\quad\times  e^{iu\hbar \o a^+a +i u L(t_{f})} (a^+-z^*)^n |
z \rangle \nonumber \allowdisplaybreaks \\
&=\frac{1}{n!} e^{iu L(t_{f})} \frac{\partial^{2n}}{\partial x^n \partial
  y^n} \nonumber \allowdisplaybreaks \\
&\quad \langle z| e^{x(a-z)} e^{iu \hbar \o a^+a} e^{y(a^+-z^*)} | z
\rangle|_{x=y=0}
\label{nz}
\end{align}
Here we have introduced the auxiliary variables $x$ and $y$ which allow
to represent the $n$th powers of shifted creation and annihilation
operators by derivatives of respective order. The scalar
function $e^{-i(xz + y z^*)}$ can be taken out of the scalar product
and the remaining operator can be brought into normal order. It then
becomes \cite{Wi}
\be
\begin{split}
e^{xa} e^{iu \hbar \o a^+ a} e^{ya^+} &= \mathcal{N}\left \{
  \exp\left [(e^{iu\hbar \o} -1) a^+a \right . \right .\\
&\quad \left . \left . + e^{iu\hbar\o} \left (x a + y
      a^+ + x y \right ) \right ] \right \}\; ,  
\end{split}
\ee{N} 
where  under the normal ordering operator $\mathcal{N}$ all
creation operators stand left of the annihilation operators. The
matrix element with respect to the coherent state $|z\rangle$ can be
read off, yielding,
\begin{widetext}
\be
\begin{split}
\langle n | e^{iu H_H(t_{f})} |n\rangle =&\: \frac{1}{n!}  e^{iu L(t_{f})} \exp
\left \{ \left ( e^{iu \hbar \o}-1 \right ) |z|^2 \right \}
\frac{\partial^{2n}}{\partial x^n \partial y^n} \exp \left \{\left 
    (e^{iu \hbar \o} -1 \right )(xz +y z^*) +e^{iu\hbar \o} xy \right
\}|_{x=y=0}\\
=&\:\frac{1}{n!}  e^{iu L(t_{f})} \exp
\left \{ \left ( e^{iu \hbar \o}-1 \right ) |z|^2 \right \} 
\frac{\partial^n}{\partial y^n} \left [ \left (e^{iu\hbar \o} -1
  \right )z + e^{iu \h \o} y \right ]^n \exp \left \{ \left ( e^{iu\hbar
      \o}-1 \right ) \left ( |z|^2 + y z^* \right ) \right \}|_{y=0}\\
=& \: e^{iu |f(t_{f})|^{2}/(\hbar \o)} \exp
\left \{ \left ( e^{iu \hbar \o} -1 \right ) |z|^2 \right \} 
\sum_{k=0}^n \binom{n}{k} \frac{|z|^{2(n-k)}}{(n-k)!}
e^{iu\hbar \o k} \left ( e^{iu\hbar \o} -1 \right )^{2(n-k)} \; .
\end{split}
\ee{nxy}
\end{widetext}
\section{Work distribution for a canonical initial state}
\label{C}
To determine the expression (\ref{qc2}) for the work
distribution $q^{\text{c}}(\tilde{\b})$ we start from 
the general expression given in the first line of 
eq. (\ref{q}). Interchanging 
the summation over the indices $n$ and $k$ we obtain
\begin{align}
q^{c}_{r}(\tilde{\b})& = e^{-|z|^{2}}\sum_{m,k=0}^{\infty} \sum_{l=0}^{2k}
(-1)^{l}\frac{|z|^{2(k+m)}}{m!\:k!} \binom{2k}{l}\allowdisplaybreaks
\nonumber \\
&\quad \times \d_{l+m,k+r}
\sum_{n=k}^{\infty} \frac{e^{-\tilde{\b} n}}{1-e^{-\tilde{\b} }}
\binom{n}{k} \allowdisplaybreaks\nonumber \\
&\stackrel{(1)}{=}e^{-|z|^{2}} \sum_{m,k=0}^{\infty} \sum_{l=0}^{2k}
(-1)^{l}\frac{|z|^{2(k+m)}}{m!\:k!} \binom{2k}{l}\nonumber \allowdisplaybreaks\\
&\quad \times  \left
  ( \frac{1}{e^{\tilde{\b} }-1}\right )^{k}
\d_{l+m,k+r}\allowdisplaybreaks \nonumber \\
&\stackrel{(2)}{=}(-1)^{r}e^{-|z|^{2}}
\sum_{m=0}^{\infty}\frac{(-|z|^{2})^{m}}{m!}\nonumber \allowdisplaybreaks\\
&\quad \times \sum_{k=|m-r|}^{\infty}  
\frac{\left (-|z|^{2}/(e^{\tilde{\b}}-1) \right )^{k}}{k!}
\binom{2k}{k+r-m}\allowdisplaybreaks\nonumber \\
&\stackrel{(3)}{=}(-1)^{r}e^{-|z|^{2}}
\sum_{m=0}^{\infty}\frac{(-|z|^{2})^{m}}{m!}
e^{-2|z|^{2}/(e^{\tilde{\b}}-1)} \nonumber \allowdisplaybreaks\\
&\quad \times I_{|m-r|}\left (- \frac{2
    |z|^{2}}{e^{\tilde{\b}}-1}\right )\nonumber \allowdisplaybreaks\\ 
& \stackrel{(4)}{=}e^{-|z|^{2} \coth (\tilde{\b}/2)}
\sum_{m=0}^{\infty} \frac{|z|^{2m}}{m!} I_{|m-r|} \left (\frac{2
  |z|^{2}}{e^{\tilde{\b}}-1} \right )\nonumber \allowdisplaybreaks\\
& \stackrel{(5)}{=} e^{-|z|^{2} \coth (\tilde{\b}/2)} \left \{ 
 \sum_{m=0}^{\infty} \frac{|z|^{2}}{m!} I_{r-m} \left (\frac{2
  |z|^{2}}{e^{\tilde{\b}}-1} \right ) \right .\nonumber\allowdisplaybreaks\\
&\quad \left . +\sum_{m=r+1}^{\infty} 
\frac{|z|^{2}}{m!}\left [I_{m-r} \left (\frac{2
  |z|^{2}}{e^{\tilde{\b}}-1} \right ) \right . \right
.\nonumber\allowdisplaybreaks\\ 
&\left . \left .  \quad -I_{r-m} \left (\frac{2
  |z|^{2}}{e^{\tilde{\b}}-1} \right ) \right] \right \}
\allowdisplaybreaks\nonumber\\ 
& 
= e^{-|z|^{2} \coth (\tilde{\b}/2)}
e^{\tilde{\b}r/2} I_{r}\left(\frac{|z|^{2}}{\sinh (\tilde{\b}/2)} \right).
\label{qcp}
\end{align}
In the first step ($\stackrel{(1)}{=}$) we  performed the sum on $n$
according to 
\be
\sum_{n=k}^{\infty} \frac{x^{k}}{1-x} \binom{n}{k} = \left (
  \frac{x}{1-x}\right)^{k},
\ee{s1} 
cf. Ref.~\cite{PBM}, 5.2.11.3. In the second step $\stackrel{(2)}{=}$
the Kronecker delta is used 
to perform the sum over k. The third step $\stackrel{(3)}{=}$ is based
on the relation 
\be
\sum_{k=|l|}^{\infty} \frac{x^{k}}{k!} \binom{2k}{k+l} = e^{2x}
I_{|l|}(2x)
\ee{s2}
valid for integer $l$. Here $I_{\n}(x)$ denotes the modified Bessel
function of the first kind of order $\n$. 
With $I_{\n}(-x) =(-1)^{\n}I_{\n}(x)$ where
$\n$ is an integer, we come to the right hand side of the equality 
$\stackrel{(4)}{=}$. In the next step the sum on $m$ is rewritten. The
term in the square brackets vanishes because $I_{\n}(x)$ is an even
function of the order $\n$. The remaining sum can be performed by
means of the identity
\be
\sum_{k=0}^{\infty} \frac{t^{k}}{k!} I_{\n -k} (x) = \left ( \frac{2
    t}{x} +1\right )^{\n/2} I_{\n} \left( \sqrt{x^{2}+2 t x}\right )\; ,
\ee{sB}
cf. \cite{PBM} 5.8.3.1.
This leads to the final 
result given in eq. (\ref{qc2}).
\section{Work distribution for a coherent initial state eq. (\ref{qcs})}
\label{D}
Starting from eq.~(\ref{q}) we may proceed in an analogous way as 
in the case of a canonical
initial state, cf. the Appendix~\ref{C}. 
According to eq. (\ref{pcs}) 
a Poissonian average over the binomial
$\binom{n}{k}$ has to be performed
instead of the geometric average in the first step of eq. (\ref{qcp}).
This yields
\be
\sum_{n=k}^{\infty} \frac{|\a|^{2}}{k!} e^{-|\a|^{2}} \binom{n}{k}
= \frac{|\a|^{2k}}{k!} \; .
\ee{qcs1}
Next the Kronnecker delta is used to perform the sum over $l$ leaving
one with two  sums of which the inner one over k can be expressed in
terms of a generalized hypergeometric function, \cite{RG}, to become
\begin{widetext}
\be
\sum_{k=|m-r|}^{\infty} \frac{(-|\a z|^{2})^{k}}{(k!)^{2}}
\binom{2k}{k+r-m} = \frac{(-|\a z|^{2})^{|m-r|}}{(|m-r|!)^{2}}\\
{}_1F_{2} \left( |m-r|+\frac{1}{2};|m-r|+1,2|m-r|+1;-4|\a
  z|^{2}\right) \; .
\ee{qcs2}
\end{widetext}
This immediately leads to the expression in eq. (\ref{qcs}).

\end{document}